# Biomimetic potassium selective nanopores

Elif Turker Acar, [1,2,#,*] Steven F. Buchsbaum,[3,#,*] Cody Combs,[1] Francesco Fornasiero,[3*]

Zuzanna S. Siwy[1,4,5,*]

[1]Department of Physics and Astronomy, University of California, Irvine

[2]Department of Chemistry, Faculty of Engineering, Istanbul University Cerrahpasa,

Avcılar-Istanbul, Turkey

[3] Lawrence Livermore National Laboratory, Livermore, CA 94550

[4]Department of Chemistry, [5]Department of Biomedical Engineering, University of

California, Irvine, CA 92697

**Abstract**: **Reproducing the exquisite ion selectivity displayed by biological ion channels in artificial nanopore systems has proven to be one of the most challenging tasks undertaken by the nanopore community, yet a successful achievement of this goal offers immense technological potential. Here we show a strategy to design solid-state nanopores that selectively transport potassium ions and show negligible conductance for sodium ions. The nanopores contain walls decorated with 4′-aminobenzo-18-crown-6 ether and ssDNA molecules located at one pore entrance. The ionic selectivity stems from facilitated transport of potassium ions in the pore region containing crown ethers, while the highly charged ssDNA plays the role of a cation filter. Achieving potassium selectivity in solid-state nanopores opens new avenues toward advanced separation processes, more efficient biosensing technologies and novel biomimetic nanopore systems.**

---

# These authors contributed equally to this work; * zsiwy@uci.edu, elifacar@istanbul.edu.tr, buchsbaum1@llnl.gov, fornasiero1@llnl.gov

Since the discovery of biological channels and their importance in physiological processes, scientists have attempted to create robust man-made structures that exhibit transport properties mimicking those of their biological counterparts.(*1-5*) Responsiveness to external stimuli and exquisite ionic selectivity are two of the most exciting properties for which efficiency remains unmatched by solid-state nanopores.(*6*) Stimuli in the form of electrical potential modulation, chemical interactions, or mechanical stress can induce so called gated channels to switch between ion conductive and closed states. Biological channels are also frequently able to differentiate between ions of the same charge so that, for example, potassium-selective channels can transport potassium ions thousands times faster than sodium ions.(*7-9*) Matching biological gating and ion selectivity capabilities in a synthetic nanopore platform could not only enable new sensing technologies but also lay the groundwork for a deeper understanding of ionic and molecular transport at the nanoscale with simple and robust model systems.

Gating has been successfully achieved in a number of man-made systems. Current rectification(*10, 11*) and voltage-responsive pore opening(*12-14*) were demonstrated in polymer and solid state nanopores having walls modified, for example, with single-stranded DNA (ssDNA), or polymer brushes. Transport properties of nanopores can also be made responsive to pressure(*15-17*) or molecules present in a solution, thus mimicking the behavior of ligand-gated channels.(*1, 18-21*)

The ability to efficiently transport one monovalent cation over another monovalent cation underpins key cellular processes such as signal propagation in neurons but is very challenging to achieve in artificial nanopores. The literature has shown a few examples of lipid-bilayer-supported synthetic constructs prepared by supramolecular self-assembly to stack crown-ether molecules on top of each other and create a channel.(*22-24*) These synthetic pores, however, display only modest cation/cation selectivities. In a similar approach, self-assembly of rigid macrocycles formed hydrophobic nanopores in a lipid bilayer, which exhibited excellent selectivity towards protons over potassium ions and no measurable conductance in NaCl or LiCl.(*25*) Due to their hydrophobic character, however, the macrocycle-based channels exhibited only very low, pS conductivities in ionic strength as high as 4 M.

To our knowledge, cation/cation selectivity has not been reproduced yet in solid-state nanopores. This type of nanopore offers the advantages of tunable geometry and surface properties and are far more robust than lipid-bilayer inserted channels. Thus, solid-state nanopores permit exploration into a wider range of electrochemical conditions and easier integration into fluidic devices.

In this manuscript, we demonstrate a solid-state nanopore that preferentially conducts potassium ions over sodium ions at concentrations up to 1 M and with selectivities far surpassing those previously reported in any other man-made nanopore platform. The mechanism of ionic selectivity is based on facilitated transport(*26*) of potassium ions through a nanopore with an opening less than 2 nm, whose walls are decorated with 18-crown-6 ether molecules. In polar solvents such as water, the crown ethers are known to selectively bind and release potassium ions quickly, allowing for their transport.(*27, 28*) We chose 4′-aminobenzo-18-crown-6 ether due to the presence of an amino group that permits easy attachment to a carboxylated surface.(*29*)

**Potassium selectivity is achieved via attachment of crown ether**

Figure 1a shows the fabrication of single nanopores and the types of chemical modifications the structures were subjected to. The pores used in this study were formed in 30 nm thick films of silicon nitride by dielectric breakdown.(*30-32*) Transport characteristics of the as-prepared pores were measured in 1 M and 100 mM solutions of KCl and NaCl. The pore walls and the membrane surfaces were then modified with triethoxysilylpropylmaleamic acid (TESPMA), which led to the attachment of carboxyl groups and the reduction of the pore diameter by a few nanometers.(*33*) Carboxylated nanopores were again tested in 1 M KCl, 100 mM KCl and NaCl solutions, and the recorded I-V curves were used to size the pore diameter by relating the pore resistance with its geometry (Supplementary Information). All reported pore diameters in this work are therefore calculated after TESPMA functionalization. Following carboxylation, nanopores were subjected to one of two modification strategies. (i) The first sub-set of nanopores (n=3) was decorated with 4′-aminobenzo-18-crown-6 using 1-ethyl-3-(3-dimethylaminopropyl) carbodiimide (EDC) coupling chemistry(*34*) (Figure 1b). (ii) The second sub-set of nanopores (n=6) was modified from one side with ssDNA oligomer, and the other side with the crown ether (Figure 1c). This asymmetric functionalization scheme was motivated by the goal of combining voltage-gated transport and cation/cation selectivity in a single synthetic pore that mimics the structure and double functionality of potassium gated channels.(*7, 35*) Reported selectivity of our pores towards potassium ions are defined here as the ratio of currents in KCl and NaCl solutions measured at 1 M, 100 mM and 10 mM concentrations with a voltage of +1 V across the membrane. The positive sign of electric potential difference corresponds to a working electrode on the side of the membrane with ssDNA; thus, for positive voltages, cations enter the pore from the DNA side and move towards the pore opening containing crown ethers.

Figure 1b shows recordings for a 1 nm wide nanopore modified with crown ether only. These include the ion selectivities before and after each modification step for all studied concentrations and the current-voltage curves in 1 M KCl and 1 M NaCl solutions. Selectivity towards potassium ions is evident only after attachment of crown ether so that the ionic current at +1 V in KCl solutions becomes at least 10 times higher than in NaCl solutions. The pore shows current rectification, thus the selectivity calculated at positive and negative voltages is different (Figure S1). Before implementing the asymmetric functionalization scheme (ii), we tested also the ionic selectivity of a pore subjected to crown modification only from one side. Current recordings for this pore (Figure S2) revealed that a partly modified pore still preferentially conducts potassium ions.

**DNA plays a role of a cation filter**

In our pore design inspired by voltage-gated ion channels, grafting of ssDNA was localized to the membrane surface and pore mouth through the selection of a 30-mer ssDNA, which is too large to diffuse inside the nanopore.(*36, 37*) We expected the high density of negative charges on the DNA to increase the cation concentrations at the pore entrance, thus causing the process of binding/releasing of ions from the crown ether to be the limiting step in the ion transport process. Note that potassium channels in a cell membrane also feature negative surface charges at one entrance, which are believed to increase local ionic concentrations and pore conductance.(*35*) The highly charged DNA functions as a cation filter, preventing anions from passing through, (*38*) but does not contribute to the pore selectivity towards potassium ions (Figure S3). Example recordings for a 1 nm wide pore subjected to ssDNA/crown ether modification is shown in Figure 1c.

**Potassium selectivity is based on facilitated transport**

We hypothesized that the selectivity observed in the two nanopores shown in Figure 1 is based on facilitated transport of $K^+$ ions,(*26*) which undergo binding/unbinding to crown ethers on the pore walls. In addition, we speculated that the pore opening after TESPMA (1 nm) is narrow enough to be fully occupied by the crown ether groups after EDC chemistry, thus preventing non-selective 'bulk transport' of sodium ions through the middle section of the pore. Pore diameter is therefore predicted to play a very important role in determining the magnitude of $K^+$ selectivity.

Toward validating our hypothesis, we considered six ssDNA/crown-ether modified nanopores as well as three nanopores that contained only crown ether. In Figure 2, we plotted the ratios of ionic currents recorded in KCl and NaCl solutions as a function of pore size determined after the carboxylation step. Clearly, $K^+$ ion selectivity decreases exponentially with the increase of the pore diameter and disappears for pore sizes larger than 3 nm. Importantly, thanks to the combined action of facilitated transport and negligible bulk transport, we could achieve $K^+$/$Na^+$ selectivities up to 84, which surpass those reported in other synthetic nanopores by about one order of magnitude.

Another striking finding is the increase of $K^+$ selectivity with salt concentration for the majority of nanopores with diameter below 2 nm, a dependence that contradicts expectations from an electrostatics-based selectivity mechanism.(*39*) In order to understand this observation, we looked in detail at the concentration dependence of sodium and potassium currents (Figure S4). Sodium currents in < 2 nm nanopores are nearly concentration independent, which suggests that the nanopores are indeed too small to allow for significant non-selective transport of sodium through the pore middle region. Potassium currents on the other hand do depend on the bulk concentration, thus the ratio of currents in potassium and sodium is lower at lower concentrations.

Figure 2 also offers comparison between pores modified only with crown ethers and those with both DNA and crown ether functionalities. Even though both types of nanopores can exhibit ratios of currents in KCl and NaCl solutions > 78, the crown-ether-only modified pores lose their selectivity more rapidly with the increase of pore size. This observation supports our prediction that DNA would act as a cation selectivity filter and/or contribute to further narrow the size of any non-selective transport pathway at the pore center. Overall, our finding suggests a larger robustness of the potassium selectivity in the presence of DNA.

Like the pores modified only with crown ether, pores functionalized with both ssDNA and crown ether displayed current rectification and different selectivities with voltage sign. The difference in $K^+$/$Na^+$ selectivity between positive and negative voltages can be explained considering that crown ethers acquire a positive charge in the presence of cations.[38] Consequently, a DNA/crown ether nanopore system features oppositely charged membrane surfaces, which can give rise to a diode behavior.(*40*) Continuum modeling with the Poisson-Nernst-Planck equations shows that for positive voltages ionic concentrations in the pore are higher, and that the electric potential decays nearly linearly across the entire pore length (Figures, S5, S6). Negative voltages on the other hand cause the formation of a depletion zone with a localized voltage drop. Thus, the resulting weak voltage gradient in the crown-ether-modified region induces a smaller selectivity.

Finally, we looked at the voltage dependence of potassium ions selectivity. All pores examined exhibited a selectivity increase at larger positive voltages, the magnitude of which is strongly dependent on pore size (Figure 3). The sensitivity of selectivity to voltage was maximized at a pore diameter of ~1 nm and decreased for both larger and smaller pores.

**Phenomenological model with voltage-dependent $k_{on}$ and $k_{off}$ help explain potassium selectivity of nanopores**.

Toward validating the claim that the pore behavior discussed thus far can be explained by crown-ether mediated transport, we developed a simple model, which neglects the presence of DNA and any surface charge effects. We assumed a pore with walls decorated by crown ethers arranged in a ring, which interact with cations through binding/unbinding events governed by $k_{on}$ and $k_{off}$ rate constants. For simplicity, we focused only on transport through a single crown-ether layer (Figure 3a). Both $k_{on}$ and $k_{off}$ constants depend exponentially on voltage, $V$, according to: $k_i = k_{i0} e^{-\frac{d_{ce} eV}{L_p k_B T}}$, $i=on, off$, where $\frac{d_{ce}}{L_p}$ corresponds to the fractional potential drop an ion experiences as it approaches and binds to a crown ether, $L_p$ is the pore length, $e$ the unit charge, and $k_B$ the Boltzmann constant.(20, 41, 42) The total current in the crown ether region can then be calculated from the total time, τ, an ion takes to pass through the pore, $\tau = \frac{1}{k_{on} C} + \frac{1}{k_{off}}$, the charge density calculated from the percent of bound cation/crown ether complexes, the ion concentration $C$, and the pore radius, $R_p$. To approximate a leakage current, the model assumes bulk transport in the center of the pore if $R_p > R_{mod}$, where $R_{mod}$ is the constant radial thickness of the crown-ether-modified region. We set the pore length $L_p$ equal to 30 nm to match the experimental system. The model was then fit to the combined experimental dataset of current *versus* voltage for the nanopore shown in Figure 1c, and selectivity *versus* pore diameter for all ssDNA/crown ether modified pores (Figure 2a) while keeping $R_{mod}$ constant at 0.25 nm. Best fit values of parameters are summarized in Table 1. Fitted $d_{ce}$ suggests a reasonable spacing between crown ethers of one to a few nm, while the binding constants, especially for K$^+$, differ significantly from those seen in a bulk system.(27) While at first unexpected, this result appears to be supported by previously reported theories that binding/unbinding kinetics of close packed macrocycles under nanoscale confinement can deviate substantially from bulk behavior.(*23, 24*)

**Table 1**. Values of parameters used to fit the experimental data of potassium selectivity and ion current values (Figures 2,3)

| $k_{off0,K}$ | $k_{on0,K}$ | $k_{off0,Na}$ | $k_{on0,Na}$ | $d_{ce}$ |
|---|---|---|---|---|
| $9.9 \times 10^8$ s$^{-1}$ | $5.1 \times 10^9$ s$^{-1}$M$^{-1}$ | $9.2 \times 10^6$ s$^{-1}$ | $1.1 \times 10^8$ s$^{-1}$M$^{-1}$ | 1.9 nm |

The model is able to successfully capture several key trends observed in the experimental data, the first of which is the diameter dependence of selectivity *versus* pore diameter (Figure 2a). As expected, the selectivity decreases for large pores due to leakage current while plateauing at a maximum for small pore diameters where the leakage current becomes negligible. Next, the model reproduces well the exponential current dependence on voltage along with reasonably accurate current magnitudes (Figure 3c,d). Our set of equations also replicate the experimentally observed weaker concentration dependence for NaCl *versus* KCl current, which we attribute to Na$^+$ transport being limited by the concentration independent $k_{off}$. Finally, the model clearly portrays how the voltage dependence of the K$^+$/Na$^+$ selectivity changes with pore diameter (Figure 3d). At small enough pore size, the leakage current drops to zero and the selectivity becomes voltage independent since both K$^+$ and Na$^+$ binding rates respond to voltage similarly. As the pore diameter gets too large, the leakage current begins to dominate and, thus, the selectivity becomes less sensitive to voltage. Overall the ability of this simple model to reproduce all observed trends in the data strongly supports the proposed facilitated transport mechanism, and this simple set of equations may prove to be useful in the design of future biomimetic pores.

**Conclusions**

In conclusion, we presented a solid-state nanopore system decorated with crown ethers, which render the pores highly selective toward potassium ions when the pore diameter is sufficiently small. We found that placement of ssDNA at one pore entrance made the system selectivity more robust by concentrating cations at one entrance. Future studies will focus on equipping such a potassium selective pore with a voltage-gated component to enable voltage-regulated pore opening.

**Acknowledgments**

This work was funded by NSF (CBET-1803262), and by The Scientific and Technological Research Council of Turkey (TUBITAK), 2219-International Postdoctoral Research Fellowship Program (App. No: 1059B191600613). Elif T. Acar also acknowledges the Istanbul University Cerrahpasa, Engineering Faculty Chemistry Department for additional financial support. Steven F. Buchsbaum and Francesco Fornasiero acknowledge support from the Laboratory Directed Research and Development Program at LLNL under project tracking code 18-LW-057. Work at LLNL was performed under the auspices of the US Department of Energy under contract DE-AC52-07NA27344. We are grateful to Prof. Mark S.P. Sansom and Prof. Stephen J. Tucker from the University of Oxford for discussions.

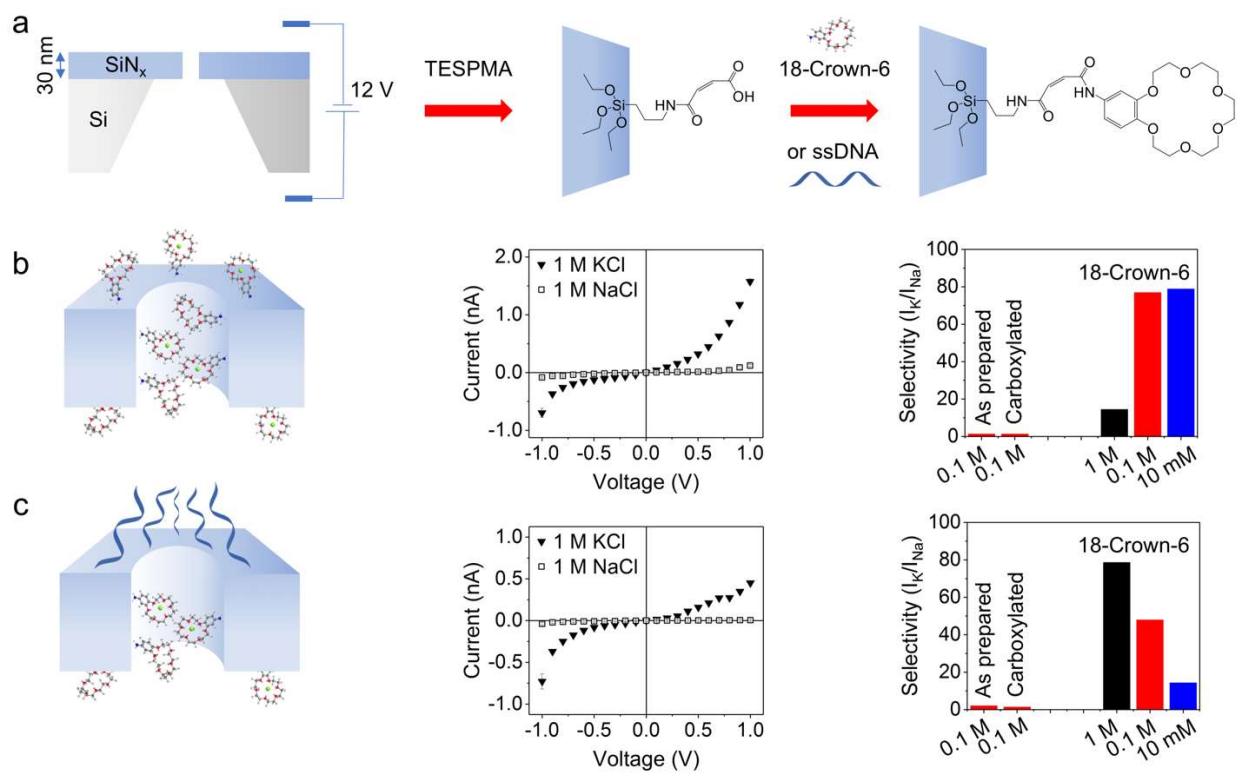

**Figure 1. Designing potassium selective solid-state nanopores**. (a) Single nanopores with tunable opening diameter were created in 30 nm thick silicon nitride films by the process of dielectric breakdown. The first modification step led to the attachment of carboxyl groups. The second modification involved either symmetric attachment of 4′-aminobenzo-18-crown-6 ether (b), or asymmetric modification with the crown ether and ssDNA (c). (b) Current-voltage curves in 1 M KCl and 1 M NaCl recorded for a 1 nm in diameter pore whose walls were decorated with crown ether, as shown in the scheme. The graph on the right summarizes ratios of currents in KCl and NaCl at 1 V before and after each modification step for the same nanopore. Ratios of currents for the nanopore before and after carboxylation are calculated based on the recordings in 100 mM of the salts. (c) Current-voltage curves in 1 M KCl and 1 M NaCl for a 0.6 nm wide

nanopore modified with crown ether and ssDNA. Selectivity of the nanopore is shown as ratios of ionic currents in KCl and NaCl solutions measured at the same conditions as in (b).

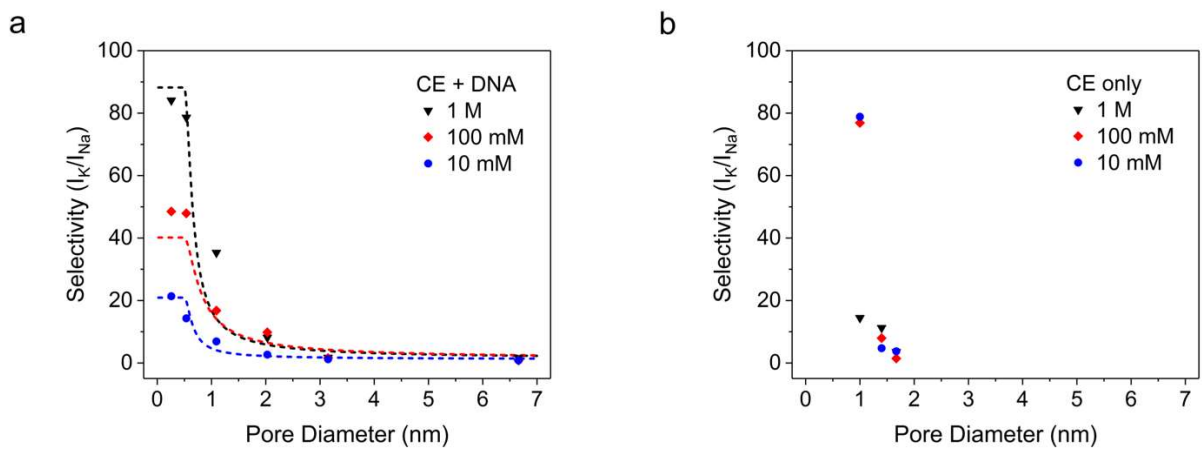

**Figure 2**. **Selectivity of nanopores towards potassium**. (a) Experimental ratios of ion currents in KCl and NaCl solutions for 6 independently prepared nanopores subjected to chemical modification with crown ethers and ssDNA. Data for three different bulk concentrations of the salts are shown. The model fit is shown as dashed lines. (b) Experimental data of potassium selectivity for three nanopores modified only with crown ether. Standard deviations of currents for individual voltages are shown in I-V curves in Figures 1, S4.

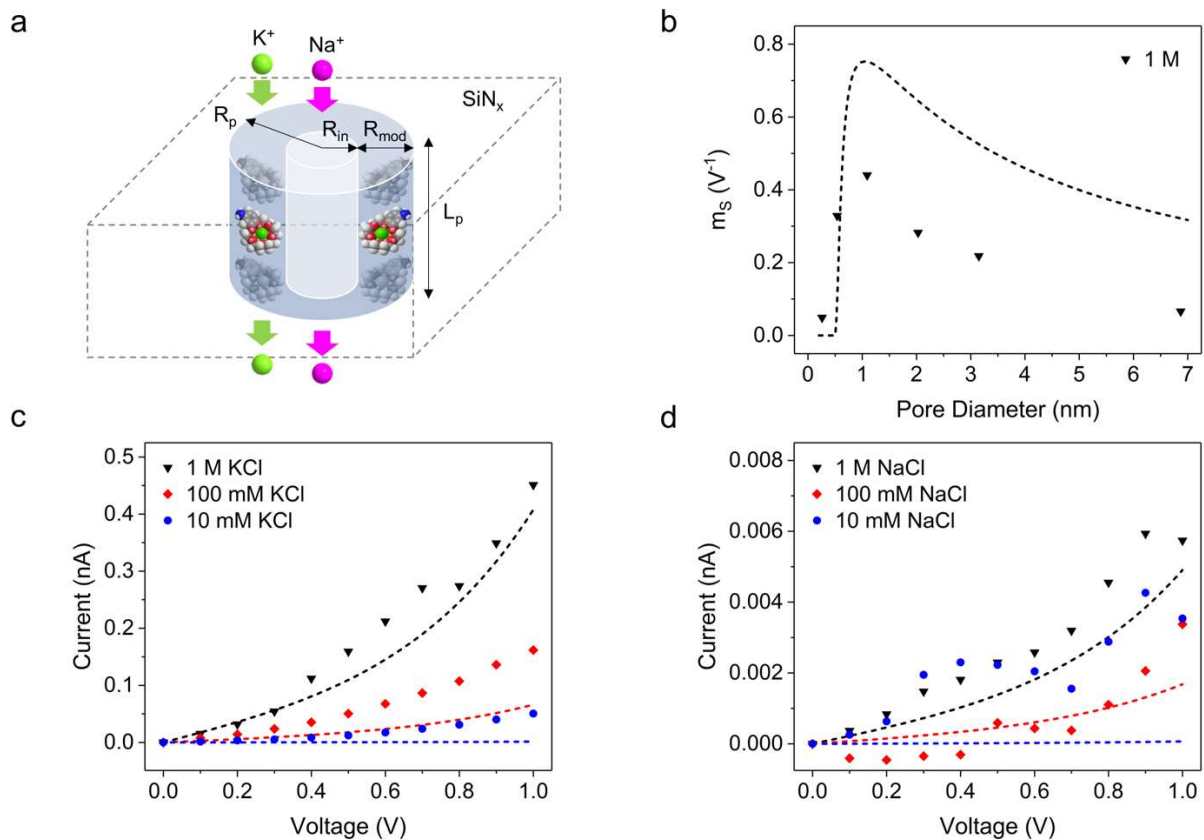

**Figure 3. Phenomenological model of the potassium selectivity of solid-state nanopores.**

(a) Scheme of the modeled system with geometrical parameters used in the model. (b) Diameter dependence of the selectivity sensitivity ($m_S$) to voltage. $m_S$ is defined here as the slope of a linear fit of Log($I_k/I_{Na}$) *versus* voltage for 1 M KCl and NaCl solution concentrations. (c) Current *versus* voltage curves at three different bulk KCl concentrations for the same pore shown in Figure 1c. (d) Current *versus* voltage curves at three different bulk NaCl concentraitons for the same pore as (c). Symbols are for experimental data, while dashed lines represent model predictions using the parameters listed in Table 1. Standard deviations of currents for individual voltages are shown in I-V curves in Figure S4.